\documentclass[11pt,a4paper,oneside]{article}
\usepackage[tight,hang]{subfigure}
\usepackage{amsmath,amssymb,amsfonts,amsthm}
\usepackage{color}
\usepackage[tight,hang]{subfigure}
\usepackage{amsmath,color,graphicx}
\usepackage{amsmath,color,graphicx}
\usepackage[english]{babel}
\usepackage{hyperref}
\usepackage{etoolbox}
\usepackage{graphicx}
\usepackage{epstopdf}
\usepackage{subfloat}
\usepackage{multirow}
\usepackage[tight,hang]{subfigure}
\usepackage{amsmath,color,graphicx}
\newcommand\overcirc[1]{\raisebox{10pt}{\tiny$\circ$}{\kern-6.5pt}\mbox{$#1$}}
\newcommand\undersym[2]{\raisebox{-5pt}{\tiny$#2$}{\kern-12pt}\mbox{$#1$}}
\newcommand\oversym[2]{\raisebox{8pt}{\tiny$#2$}{\kern-10pt}\mbox{$#1$}}
\theoremstyle{plain}

\author{Muner M. Abou Hasan$^1$, Ethar A.A. Ahmed$^2$, Ahmed F. Ghaleb$^3$,
Moustafa S. Abou-Dina$^3$, Georgios C. Georgiou$^4,*$}
\title{Numerical solution of the Newtonian plane Couette flow with linear dynamic wall slip}		

\begin{document}

\begin{center}
\Large{\bf Numerical solution of the Newtonian plane Couette flow with linear dynamic wall slip}\\

\bigskip
\large{
Muner M. Abou Hasan$^1$, Ethar A.A. Ahmed$^2$, Ahmed F. Ghaleb$^3$,
Moustafa S. Abou-Dina$^3$, Georgios C. Georgiou$^{4,*}$\\
*: Corresponding author, georgios@ucy.ac.cy
}
\end{center}


\smallskip \noindent \textit{{1: School of Mathematics and Data Sciences, EAU, Dubai, UAE.} \\[0pt]
{2: School of Engineering and Applied Sciences,
Nile University, Giza 12588, Egypt} \\[0pt]
{3: Department of Mathematics, Faculty of Science, Cairo University, Giza
12613, Egypt} \\[0pt]
{4:
Department of Mathematics and Statistics, University of
Cyprus, P.O. Box 20537, 1678 Nicosia, Cyprus}
}\newline
\vspace{3pt}
		\begin{abstract}
			An efficient numerical approach based on weighted average finite differences is used to solve
the Newtonian plane Couette flow with wall slip, obeying a dynamic slip law that generalizes the
Navier slip law with the inclusion of a relaxation term. Slip is exhibited only along the fixed plate,
and the motion is triggered by the motion of the other plate.  Three
different cases are considered for the motion of the moving plate, i.e., constant speed,
oscillating speed, and a single-period sinusoidal speed. The velocity and the volumetric flow rate
are calculated in all cases and comparisons are made with the results of other methods and available results in the literature. The numerical outcomes confirm the damping with time and the lagging effects
arising from the Navier and dynamic wall slip conditions and demonstrate the hysteretic behavior of the
slip velocity in following the harmonic boundary motion.
		\end{abstract}	
		\begin{flushleft}
		\end{flushleft}
\section{Introduction}
In these past few decades, there has been a growing interest in the study of  Newtonian and non-Newtonian viscometric flows in the presence of static or dynamic wall slip conditions, for their importance in rheometry and in industrial applications \cite{PearsonPetrie,Hat12,Hat15}. Reviews of the slip conditions prevailing at the fluid-structure interface for various media of practical importance have been reported by Hatzikiriakos \cite{Hat12,Hat15}. Malkin and Patlazan \cite{Mal-Pat18} also reviewed wall slip in complex fluids of different types, focusing on fluid-wall interaction and shear-induced fluid-to-solid transitions.

The simplest dynamic wall slip equation in the case of  unidirectional one-dimensional Newtonian flow, such that the x-velocity component is $v=v(y,t)$ and the plane  represents a wall, reads as follows \cite{PearsonPetrie}:
\begin{equation}\label{GEQ1}
	v_s(0,1)+\lambda\frac{\partial v_s}{\partial t}(0,t)=\frac{\mu}{\beta}\left| \frac{\partial v}{\partial y}(0,t)\right|,
\end{equation}
where $v_s$ is the slip velocity, defined as the relative velocity of the fluid particles adjacent to the wall with respect to that of the wall, $\mu$  is the constant fluid viscosity, $\beta$  is the slip coefficient, and $\lambda$  is the slip relaxation parameter. When the latter parameter vanishes Eq. (\ref{GEQ1}) is reduced to the Navier-slip condition \cite{Navier1827}:
\begin{equation}\label{GEQ2}
	v_s(0,t)=\frac{\mu}{\beta}\left| \frac{\partial v}{\partial y} (0,t) \right|.
\end{equation}
If the wall is fixed, then Eq. (\ref{GEQ1}) can be written as follows:
$$v(0,t)+ \lambda_s \, \frac{\partial v}{\partial t}(0,t) =\frac{\mu}{\beta}\left| \frac{\partial v}{\partial y}(0,t)\right| ,$$

Abbatiello \textit{et al.} \cite{Abb-Bul-Mar21} presented the mathematical analysis of Navier-Stokes-like problems involving a boundary where dynamic slip applies.  Ferr\'{a}s \textit{et al.} \cite{Fer-Nob-Pin12} presented analytical and semi-analytical solutions to some linear and nonlinear problems for Couette and Poiseuille flows for Newtonian and non-Newtonian media with slip boundary conditions at different walls. Thalakkottor and Mohseni \cite{Tha-Moh13} used molecular dynamic simulations to study slip at the fluid-solid boundary in an unsteady flow based on Stokes second problem, when the wall undergoes an oscillatory motion. They showed the existence of dynamic wall slip and discussed the resulting hysteresis phenomena.  Hysteresis was attributed to the unsteady inertial forces of the fluid. Kaoullas and Georgiou \cite{Kao-Geo13} and Damianou \textit{et al.} \cite{Dam-Phi-Kao-Geo14} investigated slip yield stress effects in Poiseuille flows of non-Newtonian fluids and presented analytical and numerical solutions to these problems for a variety of slip conditions. Kaoullas and Georgiou \cite{Kao-Geo15} derived analytical solutions to some Poiseuille and Couette problems including dynamic wall slip and discussed its effects on the flow development. More recent work on the flow of power law fluids in circular annuli and analytical approximate solutions was presented by Deterre \textit{et al.} \cite{Det-Nic-Lin-All-Mou20}. Pitsillou \textit{et al.} \cite{Pit-Syr-Geo20} presented solutions to flow problems with logarithmic wall slip. Ali \textit{et al.} \cite{Ali-Gha-Abo-Hel23} solved the axial, annular Couette flow of a Newtonian viscous fluid of constant density, taking into account both Navier and dynamic slip boundary conditions, using the Laplace transform technique and inversion by Laguerre polynomials.
 Farragui \textit{et al.} \cite{Far-Sou-Geo24} used separationl of variables to derive analytical solutions to the problem of cessation of annular Poiseuille  and Couette flows of a Newtonian fluid with dynamic wall slip.

The above literature review clearly shows the importance of using numerical schemes for the solution of Couette and Poiseuille flows of power-law fluids, side by side with the possibility of obtaining exact or approximate analytical solutions.

The present work uses an efficient numerical scheme to solve the Newtonian planar
Couette flow when static or dynamic wall slip applies along the fixed plate and the other plate
moves either at constant or oscillatory speed. The evolution of velocity and the volumetric flow rate are calculated for three cases: constant,
oscillating, and single-period sinusoidal plate velocity. For the last two cases, the hysteretic behavior of the fluid in following the motion of the wall is put in evidence, as this is implies energy dissipation.
\section{Governing equations}
Consider the time-dependent Couette flow of a viscous fluid between infinite parallel walls
placed a distance $d$ apart. We work with the dimensionless equations scaling lengths by $d$,
time by $d^2/ \nu$, where $\nu$ is the kinematic viscosity (defined by $ \nu \equiv /\mu/\rho$, $\rho$ being the constant fluid density), and the velocity by the characteristic
velocity $V_0$ of the moving wall. The fluid is initially at rest and suddenly the upper wall starts
moving with velocity $f(t)$, $t$ being the dimensionless time.

The governing equations and boundary conditions in dimensionless form are presented in the system (\ref{originalsys})-(\ref{originalsys2}):
	\begin{align}\label{originalsys}
	&\frac{\partial v}{\partial t}=\frac{\partial^2 v}{\partial y^2}, \quad t>0, \quad 0 \leq y \leq 1,   \\
	&v(0,t)+ \varLambda_s \, \frac{\partial v}{\partial t}(0,t) =\frac{1}{B}\frac{\partial v}{\partial y}(0,t),  \\
	&v(1,t) =  f(t), \qquad t>0 \\
	&v(y,0)=0, \qquad 0\leq y\leq1\label{originalsys2}
\end{align}
In the above equations, $v$ denotes the dimensionless velocity. No slip is assumed at the upper
wall ($y$=1) and the dynamic slip law applies at the fixed wall ($y$=0). $B=\beta d/\mu$ is the
dimensionless slip number, where $\mu$ is the viscosity, and $\varLambda_s=\lambda \nu/d^2$ is
the dimensionless slip-relaxation number. It should be noted that when $\varLambda_s=0$, the slip equation is reduced to the
static Navier slip equation. Moreover, when $B \rightarrow \infty$, the no-slip condition is
recovered.

\section{Weighted Average Finite Difference Method}
	The analytical solution of flows with dynamic wall slip is possible only for linear problems, e.g., for Newtonian flows with a linear slip equation, such as Eq. (1). However, the visualisation of the solution and the analysis of the flow, still require numerical calculations, which may not be trivial (see, e.g., \cite{S.Raju22}). Hence, it is necessary to use numerical methods to approximate the solution of this model. These approximation techniques require great effort. There exist in the literature many methods used to numerically solve partial differential equations: spectral methods \cite{Swe-Has16}, finite element, finite difference, the weighted average finite difference method (WAFDM) \cite{Swe-Gha-Abo-Has-Mek-Raw22, Swe-Has17, Swe-Ass-Has22}, and the collocation method \cite{Swe-Has19} and \cite{Swe-Has20}.
	
In what follows, we use a WAFDM scheme \cite{Smi78} to simulate and study the behavior of solutions for the problem (\ref{originalsys})-(\ref{originalsys2}) of planar Couette flow with Navier and dynamic slip conditions prevailing at the fixed wall only. The motion is triggered by assigning a given motion at the other boundary in its own plane.

The WAFDM is a widely used numerical technique for solving differential equations by approximating the derivatives of a function at discrete points in space and time using finite differences. The discretized solution is eventually  obtained by solving a system of algebraic equations \cite{Swe-Gha-Abo-Has-Mek-Raw22}. The WAFDM is relatively simple to implement, computationally efficient, and can be applied to a wide range of problems \cite{Smi78,Swe-Has16}.
The method can be  explicit (easy and simple for coding and can be used when the function being approximated is relatively smooth and well-behaved) or implicit  (more accurate and has a larger stability region and can be used when the function being approximated is not smooth), depending on the value of the weight factor $\theta$, $0 \leq \theta \leq  1$. When $\theta=0.5$,  the Crank-Nicolson implicit scheme is recovered \cite{Smi78}.

The formulation of the WAFDM is outlined below.
The domain $[0,1] \times [0, T]$ in the $(y,t)$-plane is discretized by a uniform grid with steps
$h=\Delta y$ and $k=\Delta t$, so that,
\begin{equation}
h = \frac{1}{N}, \qquad k = \frac{T}{M},
\end{equation}
where $N$ and $M$ are the numbers of subintervals used for $y$ and $t$, respectively. Hence,
the coordinates of the grid points are
%
$$y_n = n \, \Delta y, \ \ n =0, \ 1,\ 2,\ ...,\ N, \qquad t_m = m \, \Delta t,\ \ m = 0,\ 1,\ 2,\ ...,\ M.$$
The numerical values of the variable $v $ and the function $f$ at the general grid point $(y_n, t_m)$ are denoted, respectively, by $v_n^m $ and $f_n^m$. The following difference approximations are used for the time and spatial derivatives of the problem:
		\begin{align}\label{t deriv}
		(v_t)_n^m&=\frac{v_n^{m+1}-v_n^{m}}{k}+O(k),
		\end{align}
\begin{align}\label{x deriv}
	(v_y)_n^m&=\frac{v_{n+1}^{m}-v_n^{m}}{h}+O(h),
\end{align}	
and
		\begin{align}\label{xx deriv}
		(v_{yy})_n^m&=\frac{v_{n+1}^{m}-2v_{n}^{m}+v_{n-1}^{m}}{h^2}+O(h^2).
		\end{align}
These formulae are used to approximate the partial derivatives of function  $v$ in the proposed system of equations and conditions. Substituting Eqs. (\ref{t deriv})-(\ref{xx deriv}) into the governing equations (\ref{originalsys})-(\ref{originalsys2}) leads to a linear system of equations for the unknowns
$v_n^m, \  n =0, \ 1,\ 2,\dots,\ N,  m = 0,\ 1,\ 2,\dots, M$:
	\begin{align} \label{}
	&\frac{v_n^{m+1}-v_n^{m}}{k}=\theta \frac{v_{n+1}^{m}-2v_{n}^{m}+v_{n-1}^{m}}{h^2}+(1-\theta)\frac{v_{n+1}^{m+1}-2v_{n}^{m+1}+v_{n-1}^{m+1}}{h^2},\qquad \qquad \qquad \label{scheme-1} \\
	&\theta v_0^{m}+(1-\theta)v_0^{m+1}+\varLambda_s\frac{v_0^{m+1}-v_0^{m}}{k} =\frac{\theta}{B}\frac{v_1^{m}-v_0^{m}}{h}+\frac{1-\theta}{B}\frac{v_1^{m+1}-v_0^{m+1}}{h}, \label{scheme-2} \\
   & \hspace*{10cm} m=0,1,2, \dots,M \nonumber\\
& v_N^{m} = f(mk),\ \ m=1,2,3,\dots,M \label{scheme-4} \\
		&v_n^{1} =0, \ \ n=0,1,2,\dots,N. \label{scheme-5}
\end{align}

After some manipulations and simplifications, the above system may be cast in matrix form as follows:
\begin{equation} \label{mateqA}
\textbf{A} \textbf{V}^{m+1}=\textbf{B} \textbf{V}^{m}+\textbf{F}^m,
\end{equation}
where $\mathbf{V}^{m+1}$ is the vector of unknown function values at time $m+1$,
\begin{equation} \mathbf{A} =\left(
\begin{array}{c c c c c c c}
	(1-\theta)(1+\frac{1}{Bh})+\frac{\varLambda_s}{k}  \quad    & -\frac{1-\theta}{Bh} \quad  &\ \ 0  \quad  &\ \ \ 0 \quad  &\ \  \cdots\quad  &\ \ 0    \quad   & 0 \\
		 & & & & & &  \\
	a&b&a&0&\cdots&0&0 \\
0&a&b&a& \cdots&0&0 \\
0 &  0 & a & b & \cdots &0 & 0 \\
	\vdots & \vdots & \vdots & \vdots & \ddots& \vdots& \vdots \\
0 & 0 & 0& 0 & \cdots & b & a \\
	0    & 0   & 0   & 0   & \cdots & 0      &  \alpha \\
\end{array}
\right)_{N+1},
\end{equation}

\begin{equation} \label{mateqB}
\mathbf{B} =\left(
\begin{array}{c c c c c c c}
	 -\theta (1+\frac{1}{Bh})+\frac{\varLambda_s}{k}  \quad    &  \frac{ \theta}{Bh} \quad  &\ \ 0  \quad  &\ \ \ 0 \quad  &\ \  \cdots\quad  &\ \ 0    \quad   & 0 \\
		 & & & & & &  \\	
	a'&b'&a'&0&\cdots&0&0 \\
	0&a'&b'&a'& \cdots&0&0 \\
	0 &  0 & a' & b' & \cdots &0 & 0 \\
	\vdots & \vdots & \vdots & \vdots & \ddots& \vdots& \vdots \\
	0 & 0 & 0& 0 & \cdots & b' & a' \\
	0    & 0   & 0   & 0   & \cdots & 0       & 0 \\
\end{array}
\right)_{N+1},
\end{equation}
and
\begin{equation} \label{F}
\textbf{F}^m=(0, 0, 0, 0,..., f_n^m)^T_{N+1},
\end{equation}
where $f_n^m \equiv f(x_n,t_m)$,
\begin{equation}
a=-(1-\theta)\xi, \qquad b=1+ 2(1-\theta)\xi, \qquad \xi=\frac{k}{h}
\end{equation}
and
\begin{equation}
a'= \theta \xi,  \qquad b'=1-2 \theta \xi.
\end{equation}
The system \ref{mateqA} is easily solved. The local truncation error of the scheme is of order $O(h^2 + k)$.  This scheme is conditionally stable when $\theta>0.5$, and it is  unconditionally stable when $\theta \leq 0.5$ \cite{Smi78}.

\section{Results and discussion }
In this section, we use the introduced numerical scheme (\ref{scheme-1})-(\ref{scheme-5}) to simulate the approximation solution of (\ref{originalsys})-(\ref{originalsys2}). It will be shown that the introduced WAFDM provides good approximations for the solution of (\ref{originalsys})-(\ref{originalsys2}). It is applicable and efficient for solving the given system of governing equation with the accompanying boundary and initial conditions. Different test examples are carried out to approximate the solutions and compare between the solutions using different techniques.  We compared the computational time of the WAFDM and the Legendre spectral collocation method and found that the WAFDM is more efficient.

 The results assess the effect of the various parameters in the initial and boundary conditions to which the solution is subjected, demonstrate the efficiency of the introduced technique and
justify the accuracy in comparison with other methods. In our numerical experiments we choose different values for the weight factor and full  agreement is reached with the theoretical stability condition.\\

\subsection{Case 1. Plate moving at a constant speed}
\begin{figure}[htbp]
	\centering
	\includegraphics[width=0.46\textwidth]{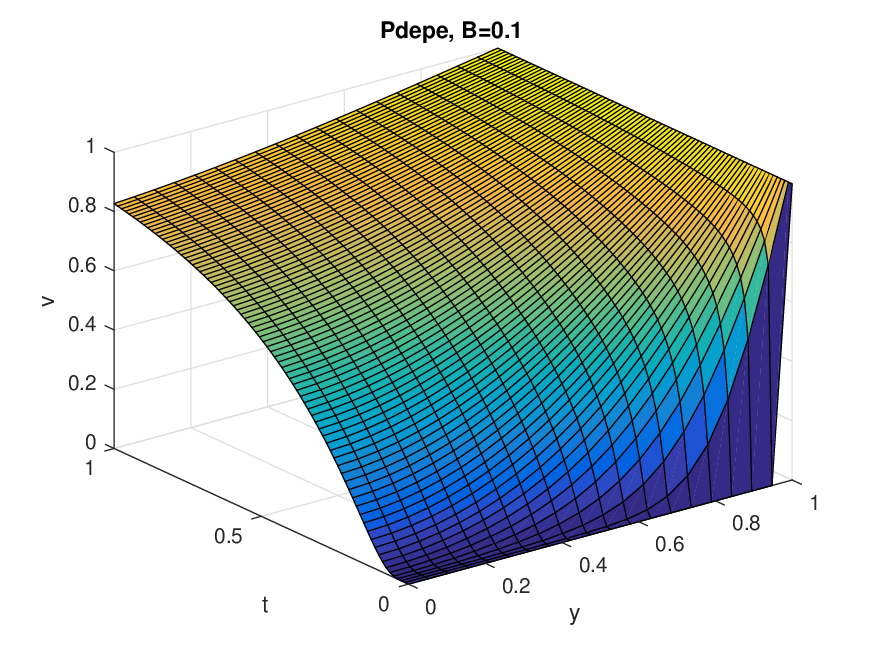}
	\includegraphics[width=0.46\textwidth]{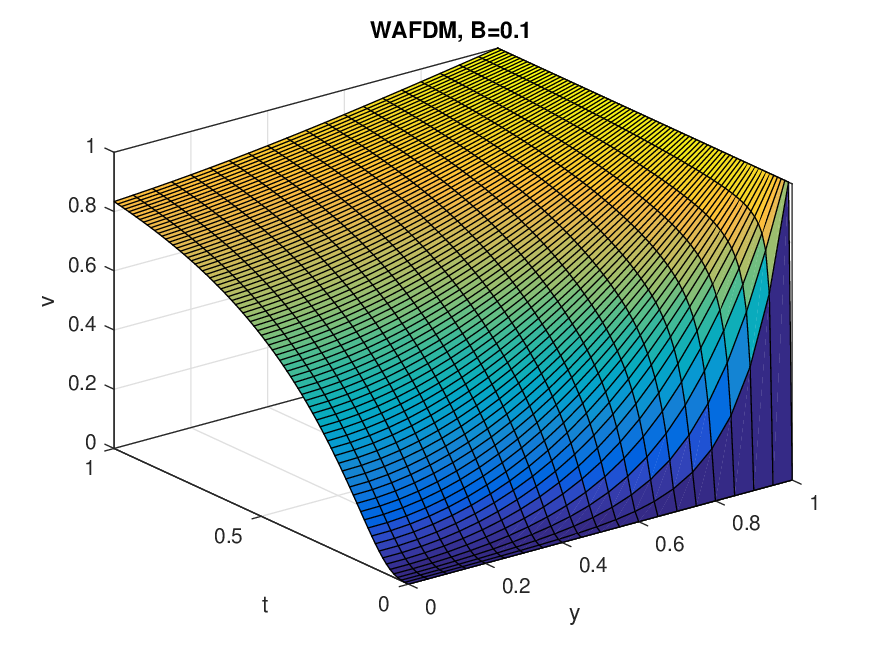}
	\includegraphics[width=0.46\textwidth]{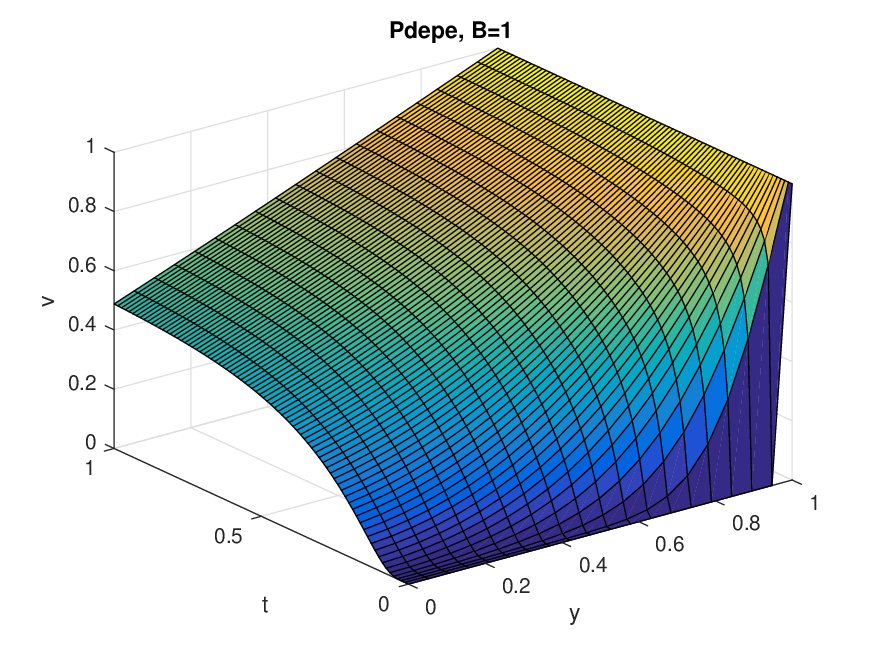}
	\includegraphics[width=0.46\textwidth]{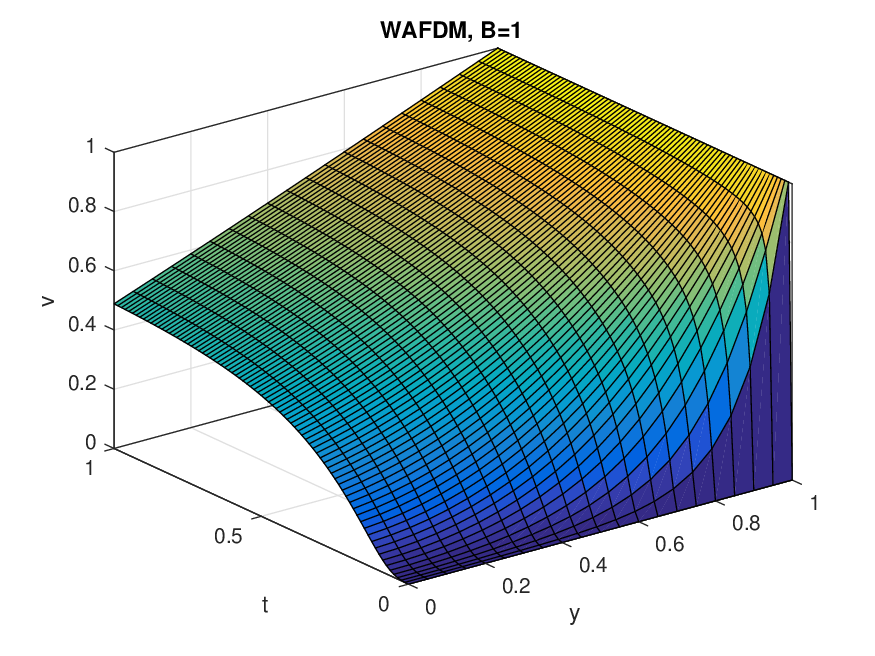}
	\includegraphics[width=0.46\textwidth]{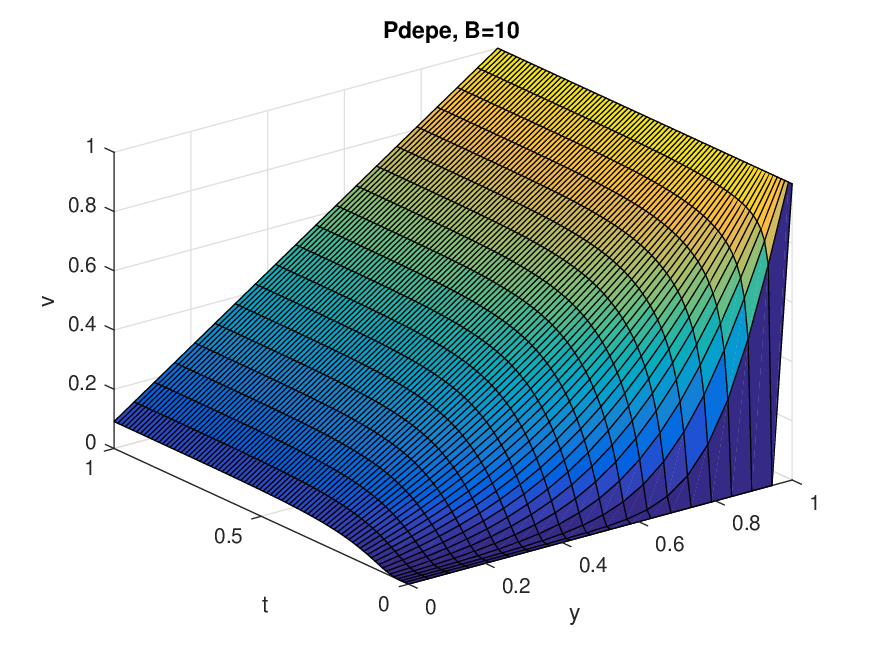}
	\includegraphics[width=0.46\textwidth]{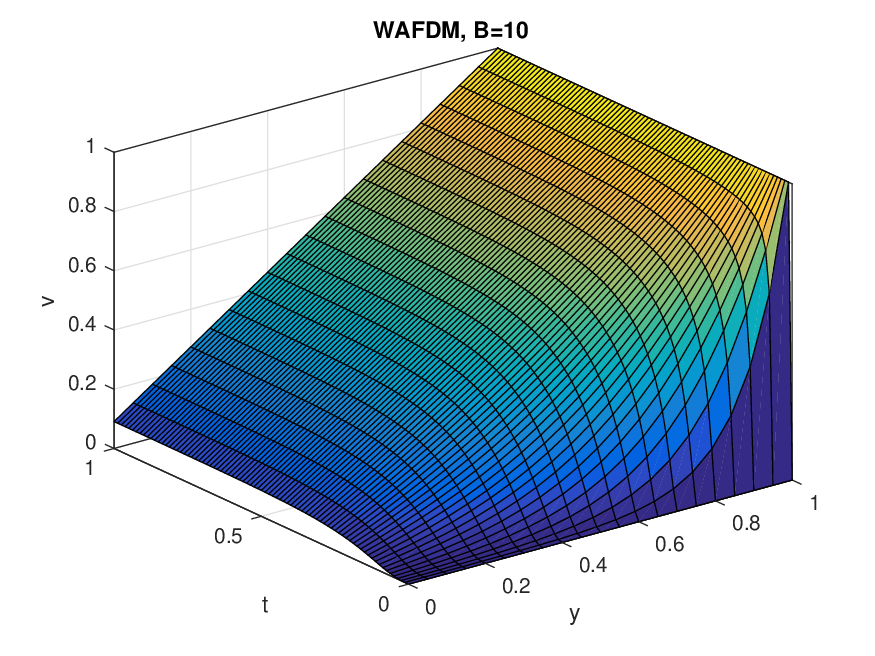}	
	\caption{Evolution of the solution $v(y,t)$  in Case 1 (plate moving at constant speed) with $\theta = 0$ and Navier slip ($ \varLambda_s=0$) for  $B=0.1$  (strong slip), 1 (moderate slip), and 10 (weak slip). The results obtained with pdepe.m (left column) compare well with those of the WAFDM (right column).
} \label{fig7}
\end{figure}

In order to assess the efficiency of the proposed numerical scheme, we have compared
our results (WAFDM) for the special case when $\theta=0$ and $f(t)=1$ with those obtained
using the Matlab pdepe toolbox. This is one of the cases treated in \cite{Abo-Hel-Gha-Kao-Geo20}, where one plate is kept
fixed, while the other one is suddenly set to motion from rest with constant speed. The results
of this comparison are shown in Figure 1 for $\varLambda_s$=0 (Navier slip) and three values of
the slip number $B$, i.e., $B$=0.1 (strong ), 1 (moderate), and 10 (weak slip).
The results of this comparison are shown in Figure \ref{fig7}. Full agreement is reached between the two methods, and with the results presented in \cite{Abo-Hel-Gha-Kao-Geo20} based on a Fourier expansion and on the use of one-sided Fourier transform as well. The main feature of the solution is the reduction of the slip velocity at the boundary $y=0$ as parameter $B$ increases.  The efficiency on the computational time of the WAFDM is demonstrated by comparing it with that of the Legendre spectral collocation method in Table 1.

\begin{table}[h!]\label{table1comparing}
	\centering
	 	\caption{Comparison of the WAFDM with the spectral collocation method. }
	\label{table1financ}
	\begin{tabular}{|l|l|l|l|l|}
		\hline
		\multicolumn{2}{|l|}{\ \ \ WAFDM} & \multicolumn{2}{l|}{\ \ \ Spectral method} \\ \hline
		$N_n$     & CPU       & $M$     & CPU     \\ \hline
		500   	  & 5.9 s    & 5     & 963.6 s  \\ \hline
		5000	  & 18.3 s  & 12     & 7579.9 s \\ \hline
		10000	  & 44.8 s  & 15     & 88937.5 s \\ \hline
		50000     & 789.9 s & 20     & 219357.6 s \\ \hline
	\end{tabular}
\end{table}

\begin{figure}[htbp]
	\centering
\includegraphics[width=0.7\textwidth]{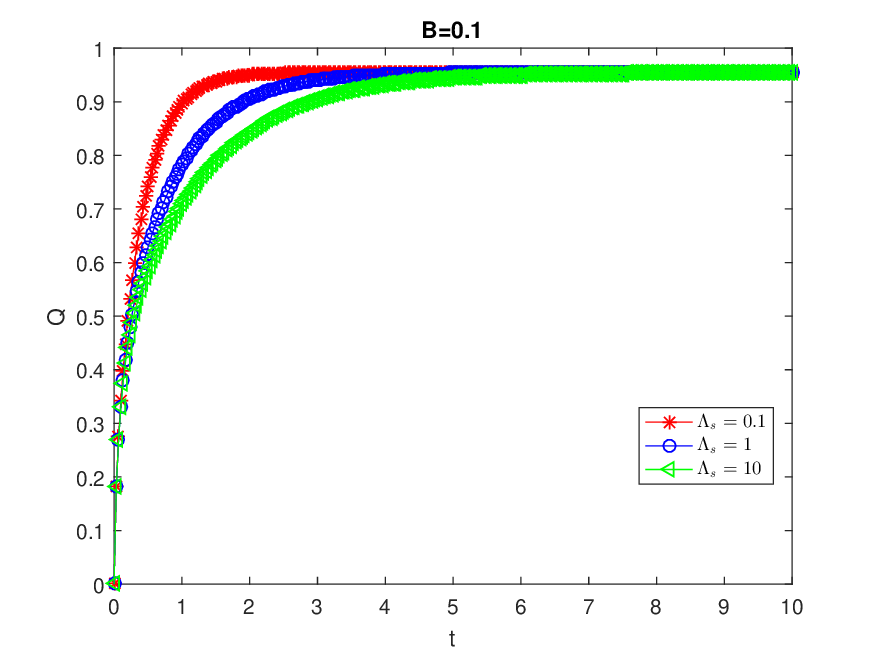}

{\bf (a)}

\includegraphics[width=0.7\textwidth]{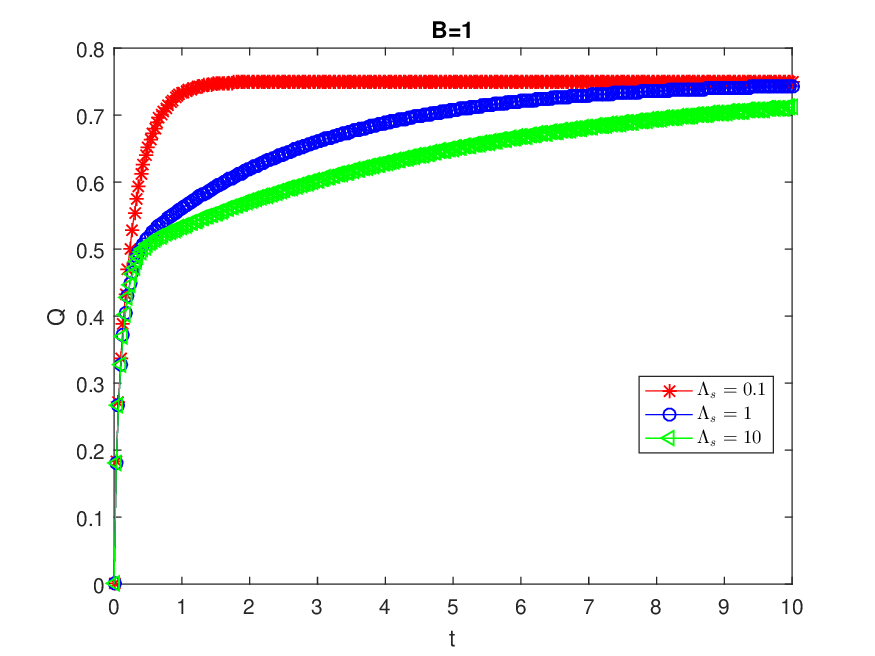}

{\bf (b)}

	\caption{Evolution of the volumetric flow rate $Q(t)$
in Case 1 (plate moving at constant speed) with  $\theta =0$ and $\varLambda_s=0$  (Navier slip), 1, and 10: (a)  $B=0.1$ (strong slip); (b) 1 (moderate slip).}\label{fig202}
\end{figure}

The volumetric flow rate, defined by \begin{equation}
	Q(t)=\int_{0}^{1}v(y,t)dy,
\end{equation} is
shown in Figure \ref{fig202} for $\theta=0$, $B$=0.1 (strong slip) and 1 (moderate slip), and  $\varLambda_s$=0 (Navier slip), 1, and 10.
The damping effect of the slip relaxation  parameter in reaching a fully-developed flow is clearly shown. As expected, the initial value of the volumetric flow rate is independent of  $\varLambda_s$ and decreases with the slip parameter $B$ (since wall slip is reduced.

\begin{figure}[htbp]
	\centering
	\includegraphics[width=0.6\textwidth]{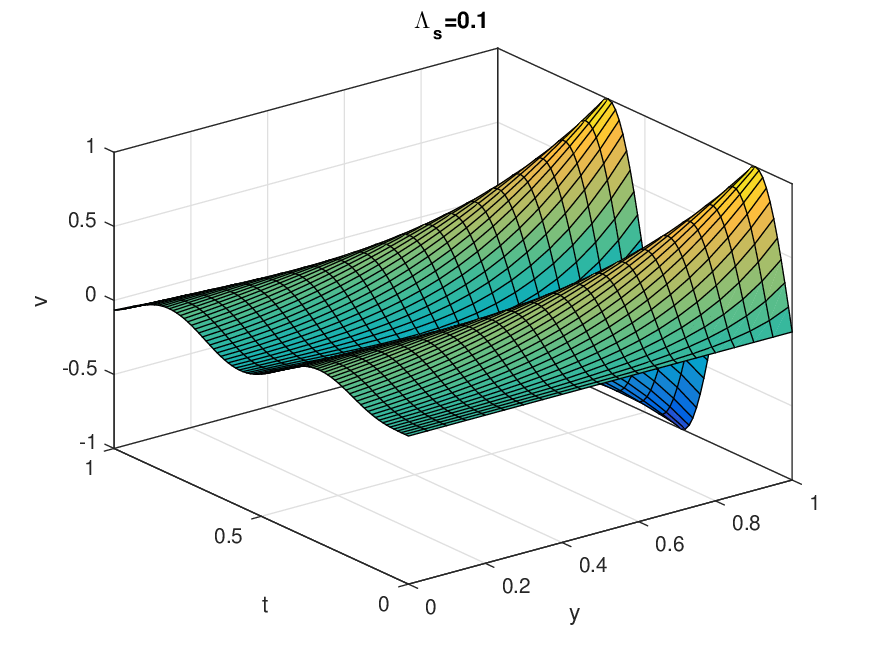}

{\bf (a)}

	\includegraphics[width=0.6\textwidth]{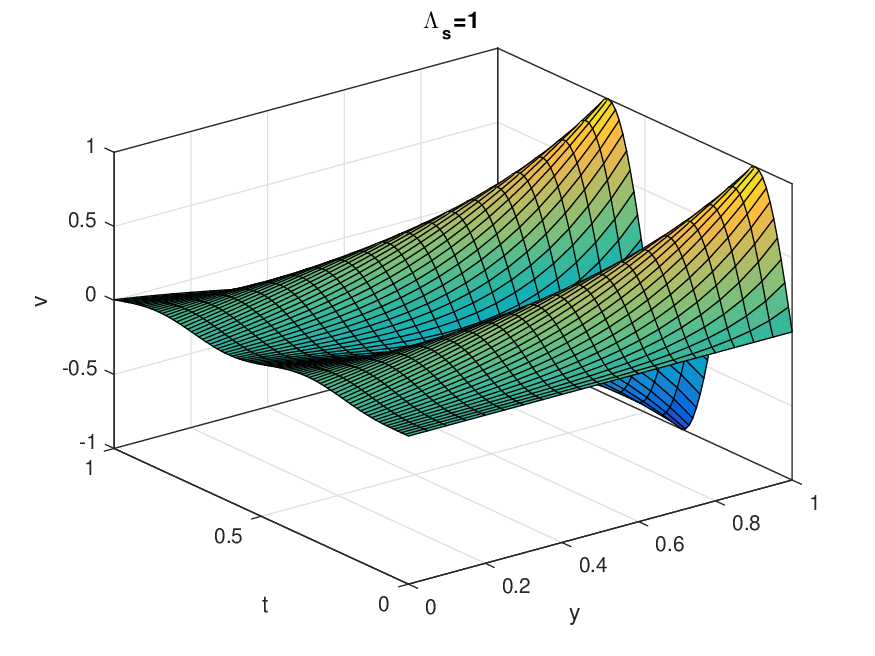}

{\bf (b)}

	\includegraphics[width=0.6\textwidth]{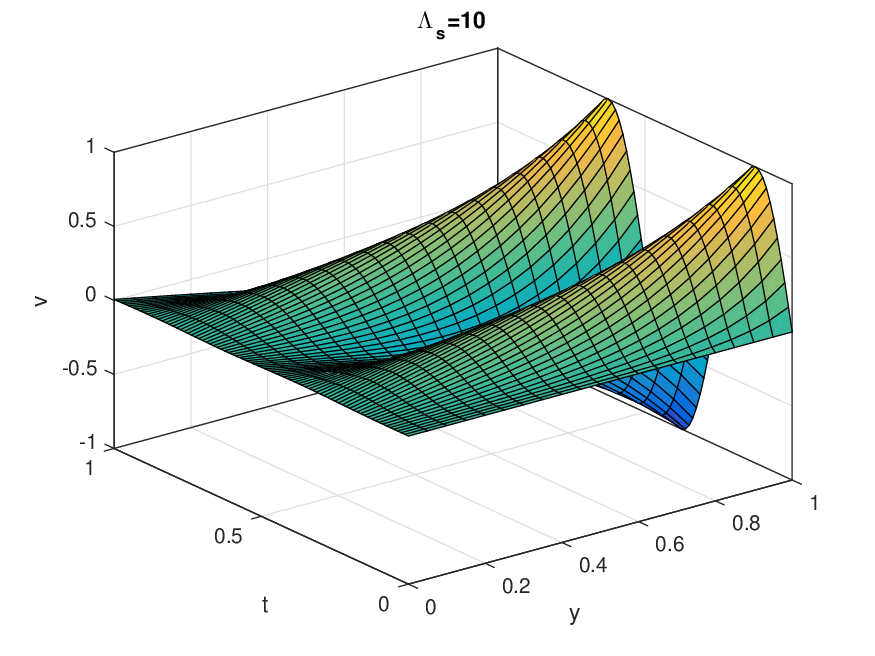}

{\bf (c)}
	\caption{Evolution of the solution $v(y,t)$ in Case 2 (oscillating plate) with $\theta =0$ and  $B=1$  (moderate slip): (a)  $\varLambda_s=0$ (Navier slip); (b) $\varLambda_s=1$; (c) $\varLambda_s=10$.
} \label{fig1}
\end{figure}

\subsection{Case 2. Oscillating plate}
Consider now the case where the moving plate is performing continuous harmonic oscillations
with period equal to unity, i.e.,
\begin{equation}
	f(t)=\sin {4 \pi t}.
\end{equation}
A similar problem was treated in \cite{Tha-Moh13}	for the unsteady flow to estimate the increment in slip at the boundary due to wall acceleration, and uncover the hysteretic behavior of the slip velocity for this harmonic motion of the boundary. Here it is required to evaluate the effect of the dynamic slip parameter $\varLambda_s$ on the flow in general, and on the hysteretic behavior of the slip velocity due to oscillating wall motion.

Figure \ref{fig1} illustrates the distribution of the solution in space and time with different values of $\varLambda_s$, as time is taken to run along two complete periods of the wall oscillations. The boundary condition at the moving wall is clearly satisfied. As $\varLambda_s$ increases from  $0.1$ to $10$, a damping of the amplitude of the oscillations with time takes place. The differences become more noticeable as one approaches the fixed wall at $y=0$, where the velocity profile becomes more flattened.

\begin{figure}[htbp]
	\centering
	\includegraphics[width=0.70\textwidth]{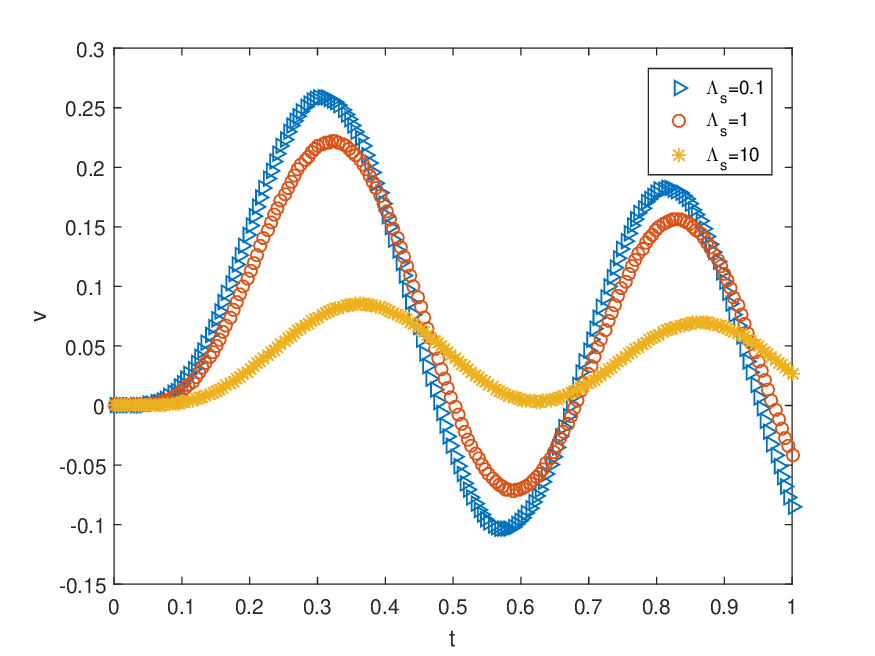}
	\caption{Evolution of the slip velocity  $v(0,t)$ in Case 2 (oscillating plate) with $\theta =0.5$  and $B=1$ (moderate slip) for  $\varLambda_s=0.1$ , 1, and 10. Time damping and lagging are observed.} \label{fig11}
\end{figure}

Figure \ref{fig11} illustrates the solutions at $y=0$, i.e the slip velocity,  as functions of time for different values of $\varLambda_s$.  The figure clearly shows time damping and lagging effects at the boundary $y=0$. The amplitudes of oscillations of the slip velocity shows a reduction of approximately $33 \%$ as the value of $\varLambda_s$ increases from $0.1$ to $10$.

\begin{figure}[htbp]
	\centering
 	\includegraphics[width=0.7\textwidth]{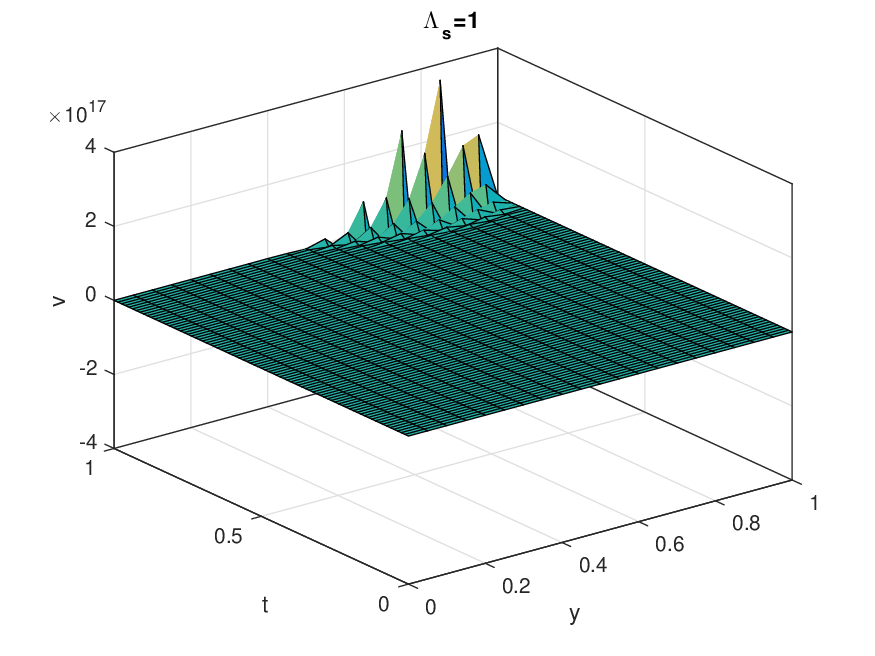}
 	\caption{Instability of the solution $v(y,t)$ in Case 2 (oscillating plate) with $\theta=0.7$, $B=1$ (moderate slip) and $\varLambda_s=1$.}\label{fig2}
\end{figure}

In order to put in evidence the crucial role played by the weight factor $\theta$ in the present numerical study, we have considered one case with a new value of this parameter to test the stability of the numerical scheme. As mentioned above, the scheme is stable provided that  $\theta < 0.5$ \cite{Smi78}.
 The instability occuring when $\theta=0.7>0.5$ is illustrated in the 3D-plot of Figure \ref{fig2}.

\begin{figure}[htbp]
	\centering
	\includegraphics[width=0.6\textwidth]{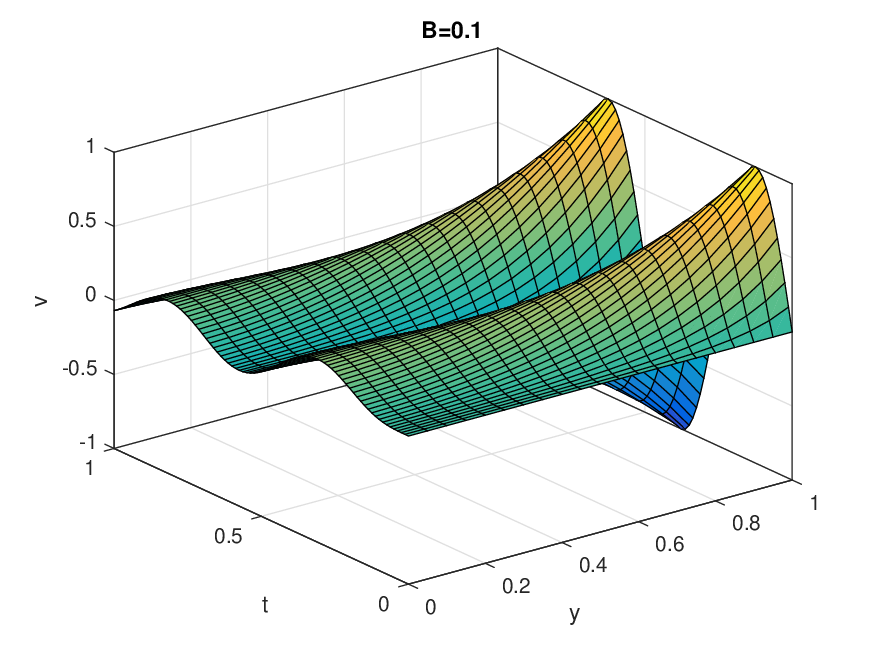}

{\bf (a)}

	\includegraphics[width=0.6\textwidth]{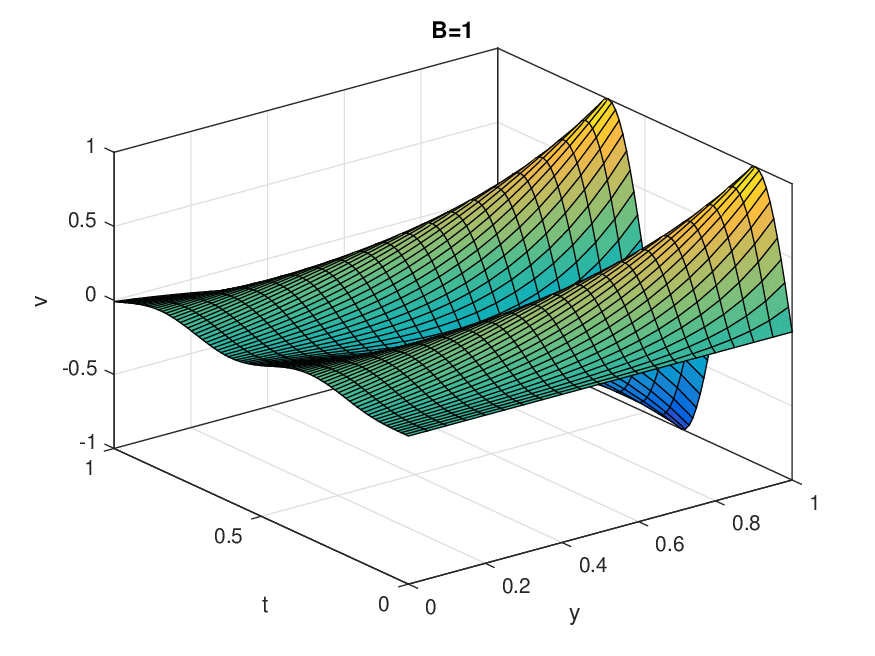}

{\bf (b)}

	\includegraphics[width=0.6 \textwidth]{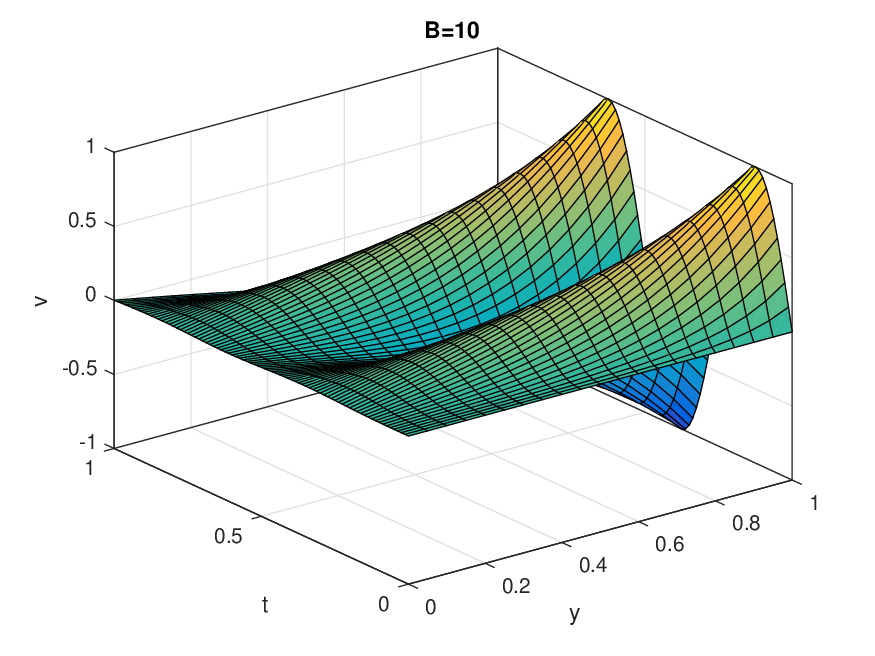}

{\bf (c)}

	\caption{Evolution of the solution $v(y,t)$ in Case 2 (oscillating plate) with $\theta =0$ and $\varLambda_s = 0.5$: (a) $B=0.1$ (strong slip); (b) $B=1$ (moderate slip); (c) $B=10$ (weak slip).} \label{fig4}
\end{figure}

\begin{figure}[htbp]
	\centering
	\includegraphics[width=0.70\textwidth]{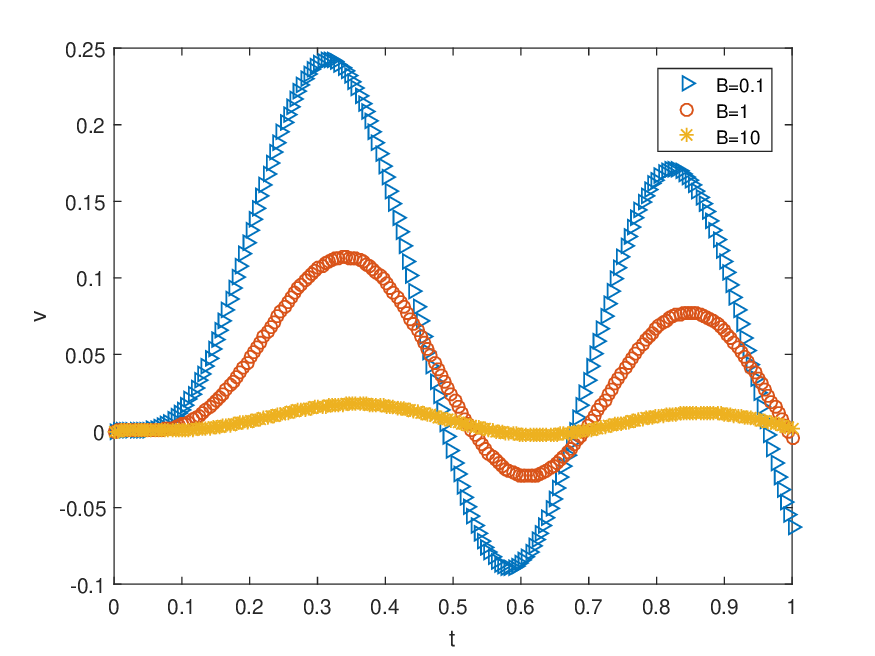}
	\caption{Evolution of the slip velocity  $v(0,t)$ in Case 2 (oscillating plate) with $\theta =0.5$ and $\varLambda_s = 0.5$ for $B=0.1$ (strong slip), 1 (moderate slip), and 10 (weak slip).} \label{fig9}
\end{figure}

The 3D plots in Figure \ref{fig4} show how the solutions change in space and time when the  slip parameter $B$ assumes the values 0.1, 1, and 10.
Figure \ref{fig9} illustrates the dependence of the slip velocity  on parameter $B$. In both plots, we have fixed the value $\varLambda_s = 0.5$. The effect of parameter $B$ becomes more pronounced as the boundary $y=0$ is approached. Here again, one notes the presence of lag in following the boundary motion, and time damping of the peaks at the boundary $y=0$ as the value of $B$ increases.

\begin{figure}[htbp]
	\centering
	\includegraphics[width=0.6\textwidth]{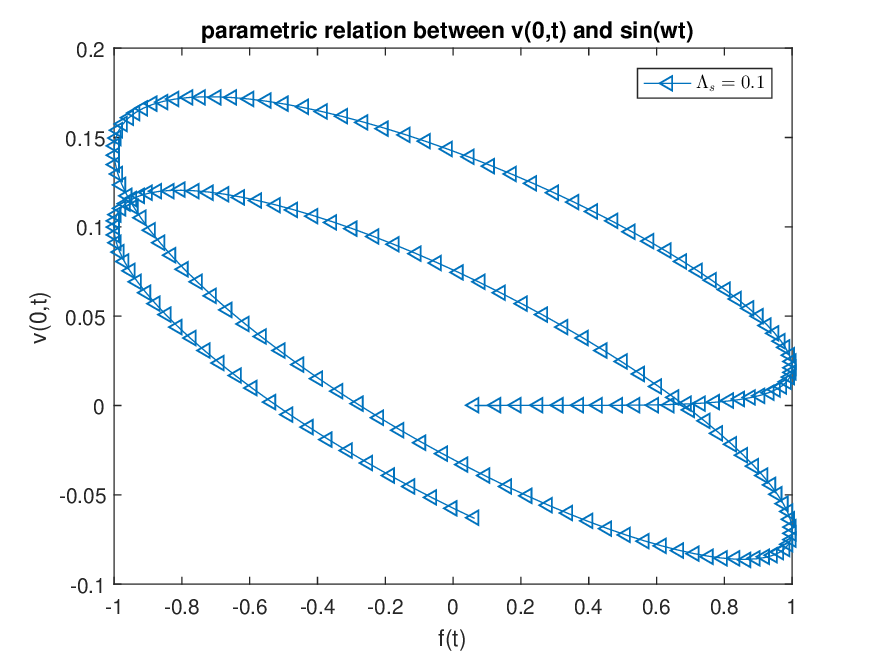}

{\bf (a)}

	\includegraphics[width=0.6\textwidth]{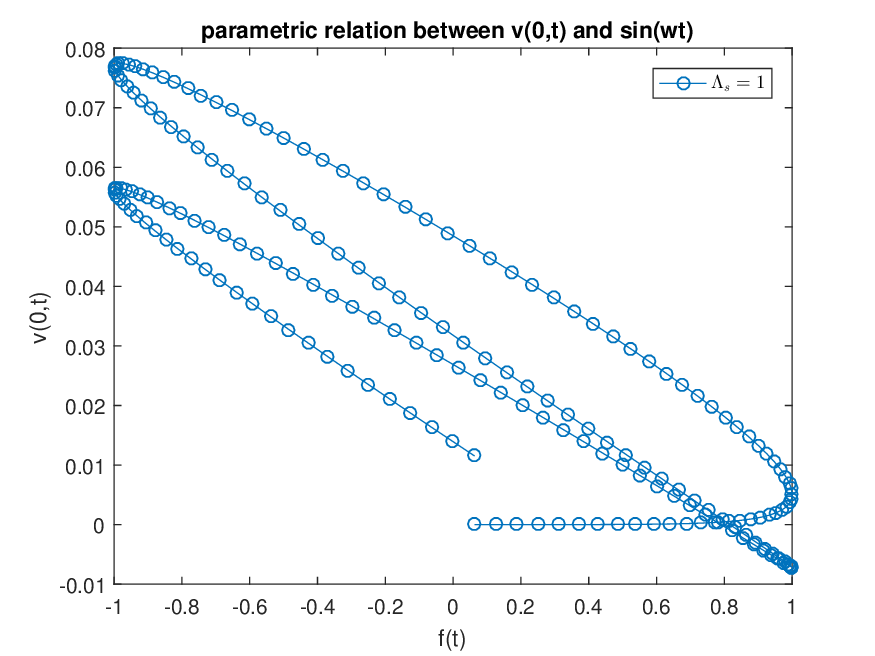}

{\bf (b)}

	\includegraphics[width=0.6\textwidth]{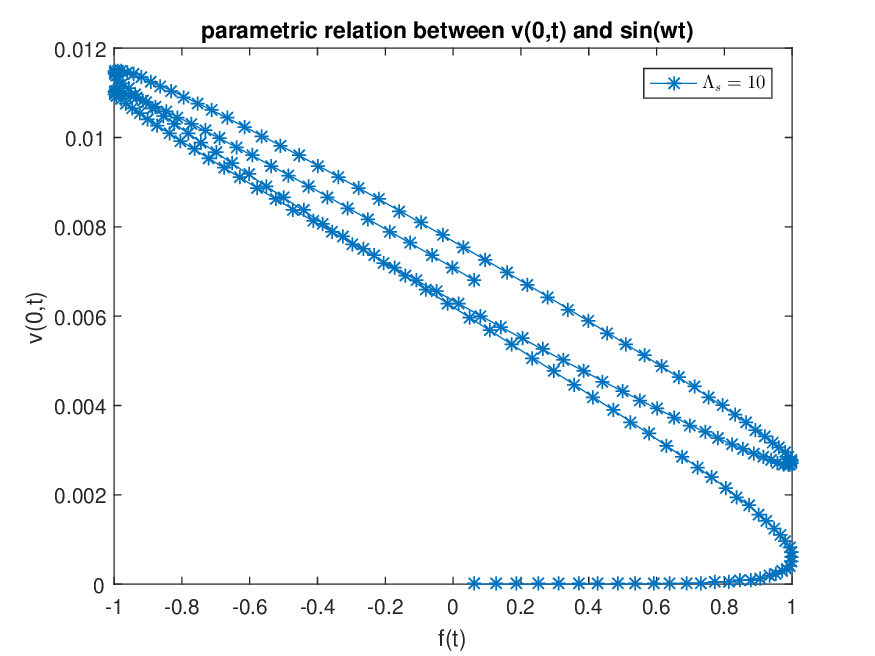}

{\bf (c)}

	\caption{Hysteretic behavior of the slip velocity $v(0,t)$ in Case 2 (oscillating plate) with  $\theta =0$ and $B=1$: (a) $\varLambda_s=0.1$; (b) $\varLambda_s=1$; (c) $\varLambda_s=10$. Note that the scale of the vertical axis changes.} \label{fig:hyst}
\end{figure}

In order to visualize the hysteretic behavior of the slip velocity and the lagging in following the applied motion on one wall noticed in \cite{Tha-Moh13}, plots  have been provided in Figure \ref{fig:hyst} of the slip velocity against the boundary motion $\sin \omega t$ for three values of the dynamic slip parameter $\varLambda_s$.  It is noticed that the hysteresis loops become narrower as the value of the dynamic slip parameter increases. In \cite{Tha-Moh13} the width of the hysteresis loop is related to
the loss of energy transfer from the wall to the fluid.

\begin{figure}[htbp]
	\centering
	\includegraphics[width=0.7\textwidth]{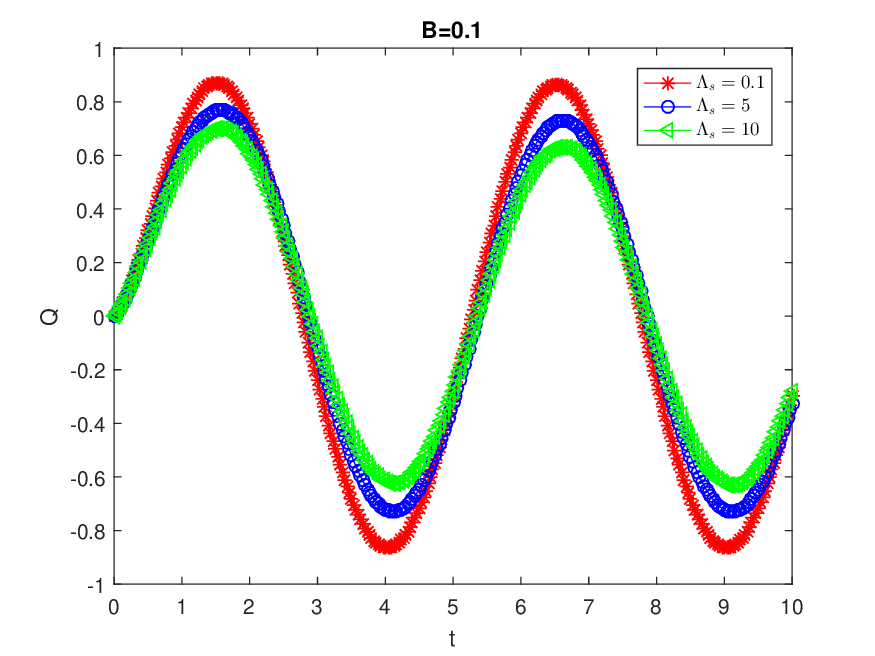}

{\bf (a)}

	\includegraphics[width=0.7\textwidth]{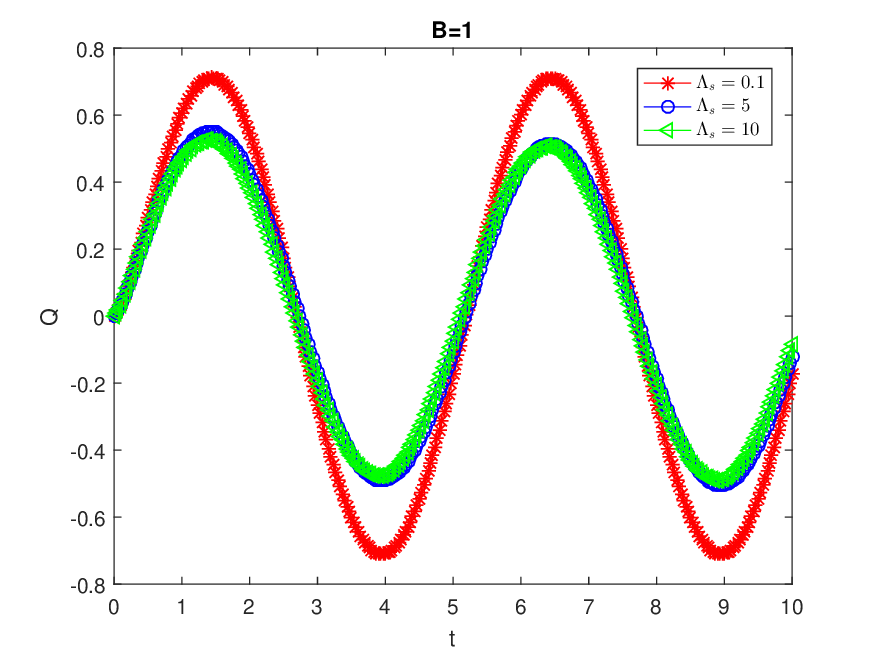}

{\bf (b)}

	\caption{Evolution of the volumetric flow rate $Q(t)$ in Case 2 (oscillating plate) with $\theta =0$ and $\varLambda_s=0.1$, 5, and 10: (a) $B=0.1$ (strong slip); (b)  $B=1$ (moderate slip).} \label{fig:a222}
\end{figure}

The volumetric flow rate is shown in Figure \ref{fig:a222}, where it is seen that this is oscillatory and is damped by the slip relaxation parameter.  However, for sufficiently large values of $\varLambda_s$, it is seen that the amplitude of oscillations of $Q(t)$ will be less sensitive to any further increase of $\varLambda_s$. This phenomenon becomes more striking for larger values of the slip parameter $B$. Notice that after two complete oscillations, the volumetric flow .flow rate has not vanished due to retardation.

\subsection{Case 3. Single plate oscillation }
In this section we assume that the upper plate oscillates only once and then comes to rest, i.e.,
\begin{equation}
f(t)=\left\{
\begin{array}{ll}
	\sin 4\pi t,& t\leq0.5\\
	0, & t>0.5
\end{array}
\right. .
\end{equation}

\begin{figure}[htbp]
	\centering
	\includegraphics[width=0.6\textwidth]{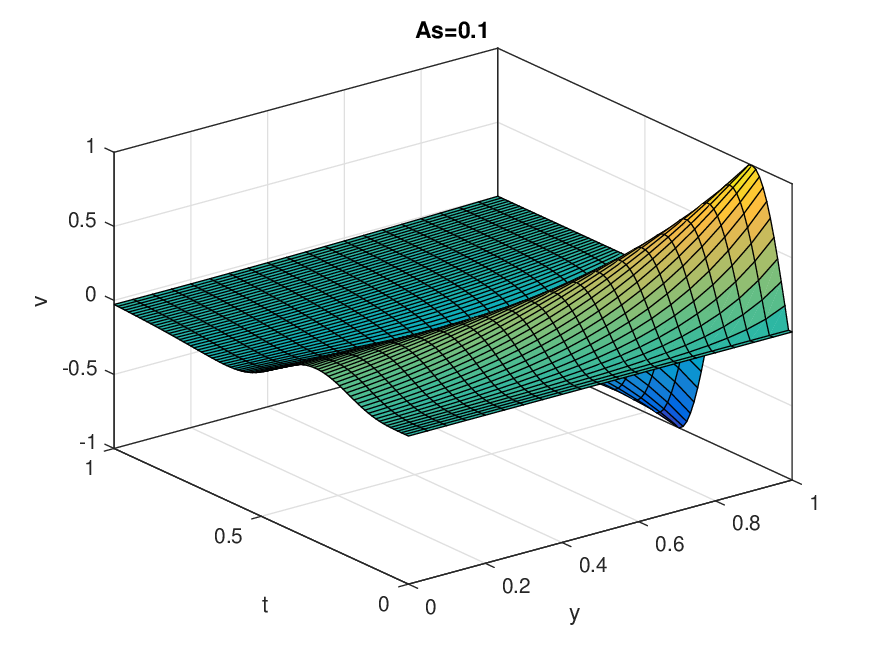}

{\bf (a)}

	\includegraphics[width=0.6\textwidth]{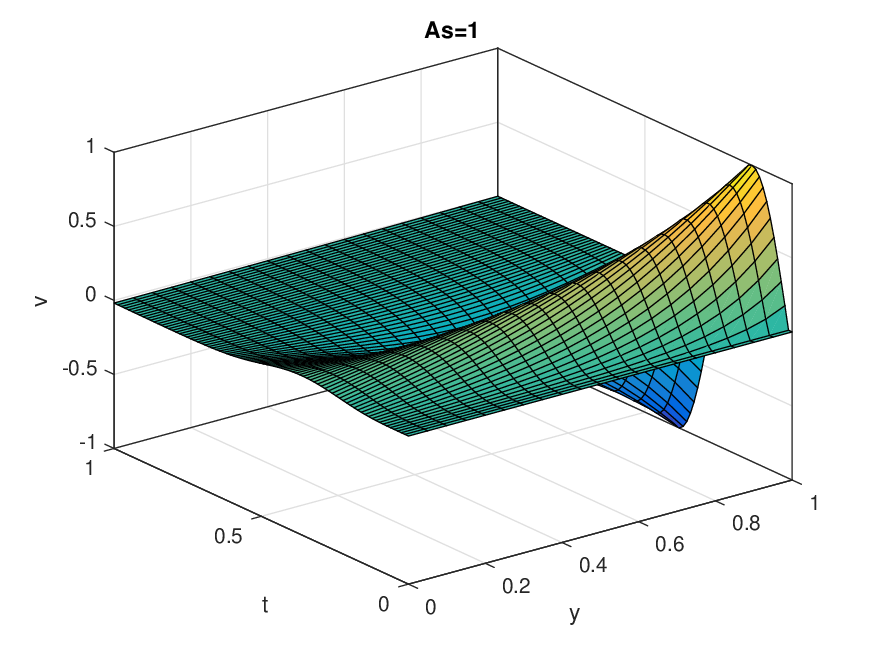}

{\bf (b)}

	\includegraphics[width=0.6 \textwidth]{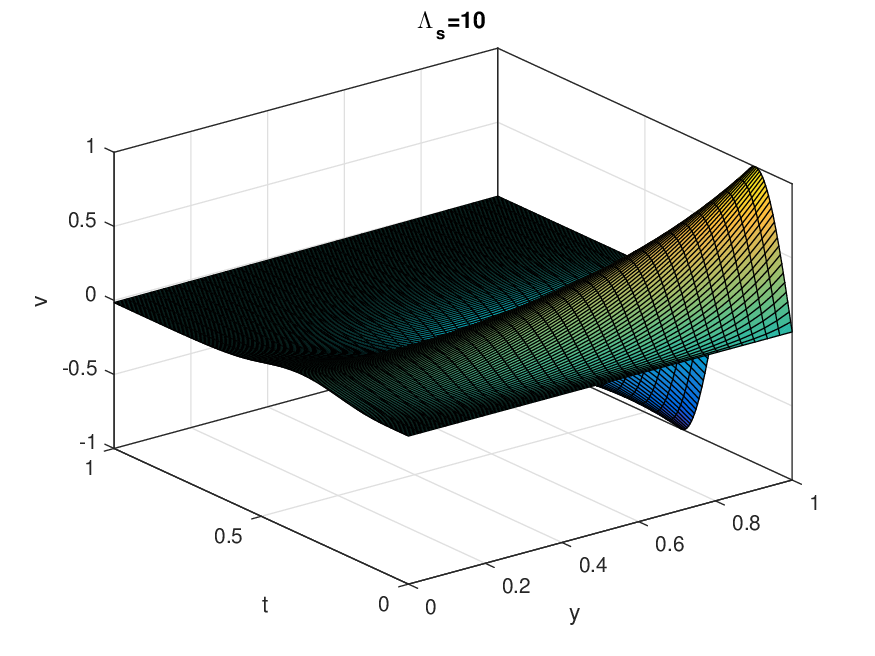}

{\bf (c)}

	\caption{Evolution of the solution $v(y,t)$ in Case 3 (single plate oscillation) with $\theta =0$ and $B=1$  (moderate slip): (a)  $\varLambda_s=0.1$; $\varLambda_s=1$; (c) $\varLambda_s=10$.}\label{fig:a3}
\end{figure}	

Figure \ref{fig:a3} illustrates the effect of the dynamic slip parameter $\varLambda_s$ on the solution, where the value of the  slip parameter was set to $B=1$.

\begin{figure}[htbp]
	\centering
	\includegraphics[width=0.78\textwidth]{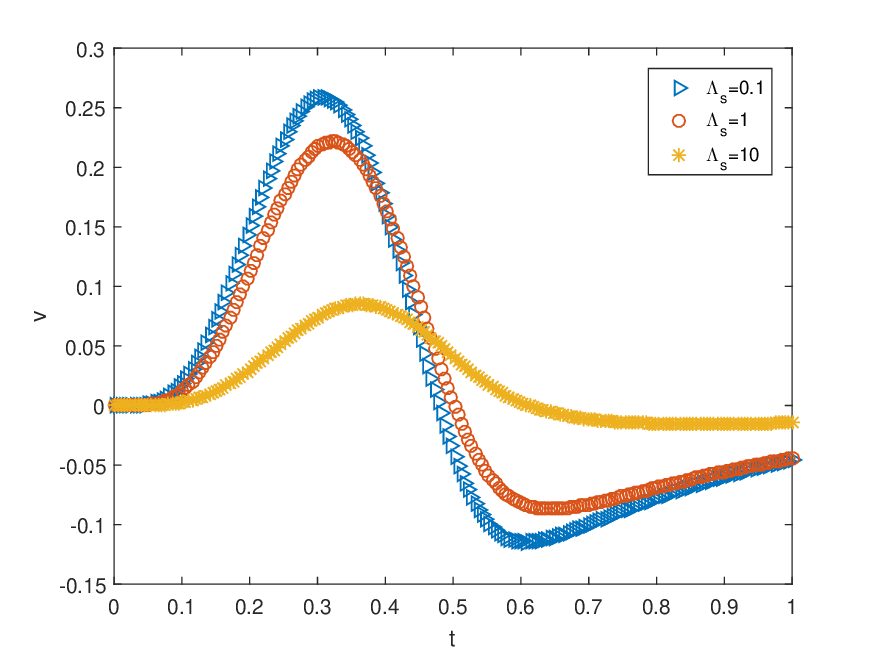}
	\caption{Evolution of the slip velocity $v(0,t)$ in Case 3 (single plate oscillation) with $\theta =0.5$ and $B=1$  (moderate slip) for $\varLambda_s=0.1$, 1, and 10.}\label{fig:a10}
\end{figure}

\begin{figure}[htbp]
	\centering
	\includegraphics[width=0.6\textwidth]{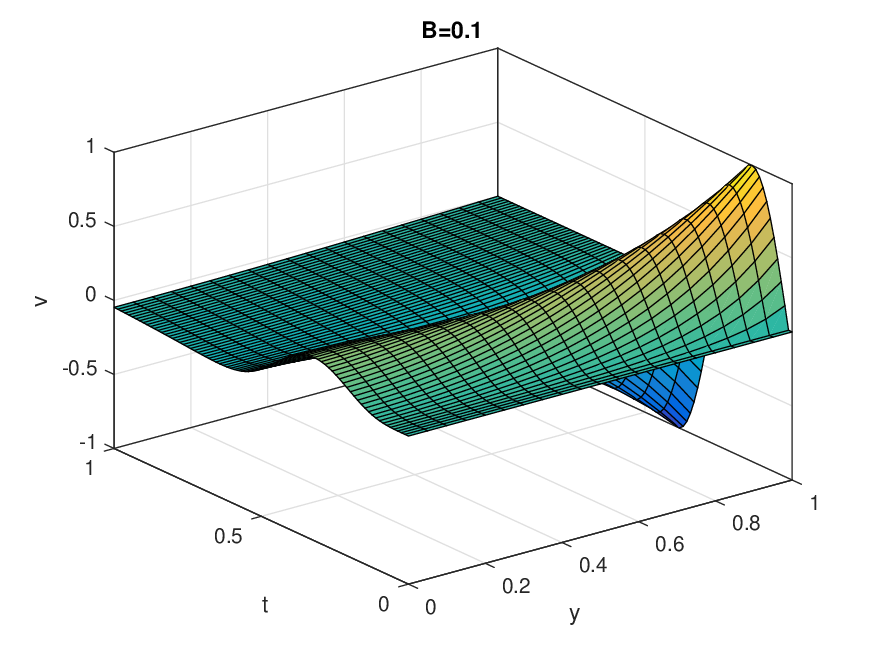}

{\bf (a)}

	\includegraphics[width=0.6\textwidth]{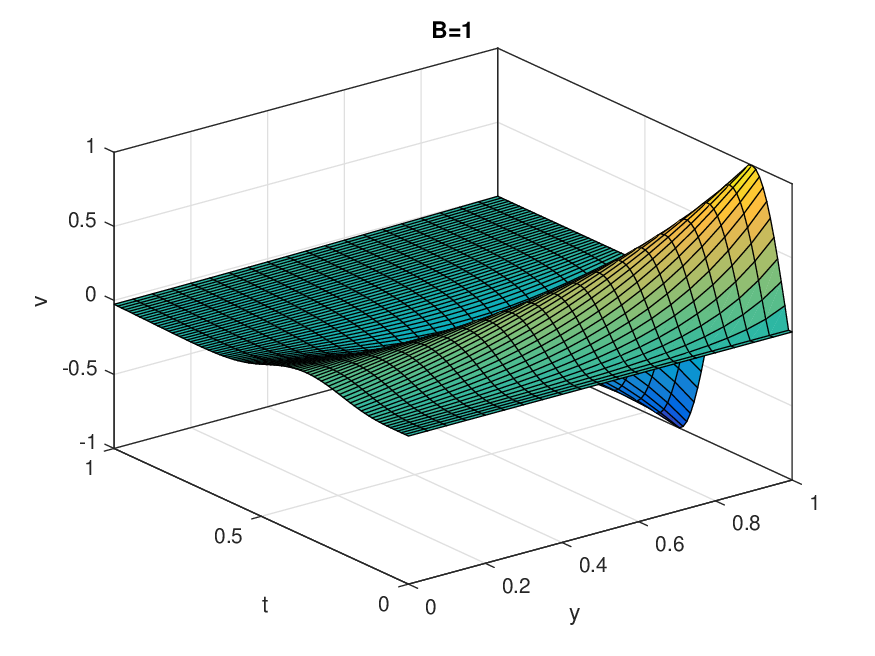}

{\bf (b)}

	\includegraphics[width=0.6\textwidth]{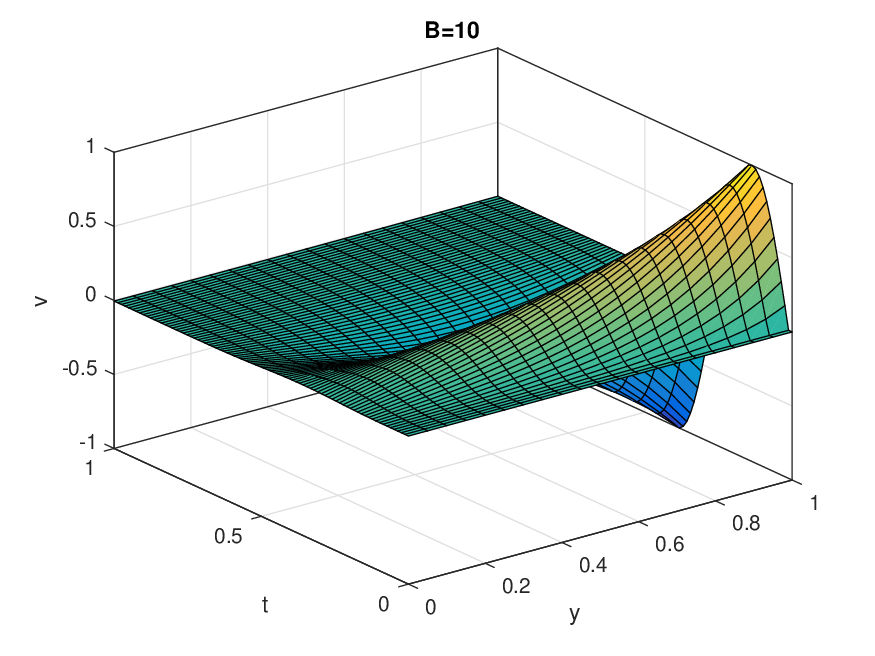}

{\bf (c)}

	\caption{Evolution of the solution $v(y,t)$ in Case 3 (single plate oscillation) with   $\theta =0$ and $\varLambda_s=0.5$: (a) $B=0.1$ (strong slip); (b) $B=1$ (moderate slip); (c) $B=10$ (weak slip).}\label{fig:a5}
\end{figure}

\begin{figure}[htbp]
 	\centering
 	\includegraphics[width=0.78\textwidth]{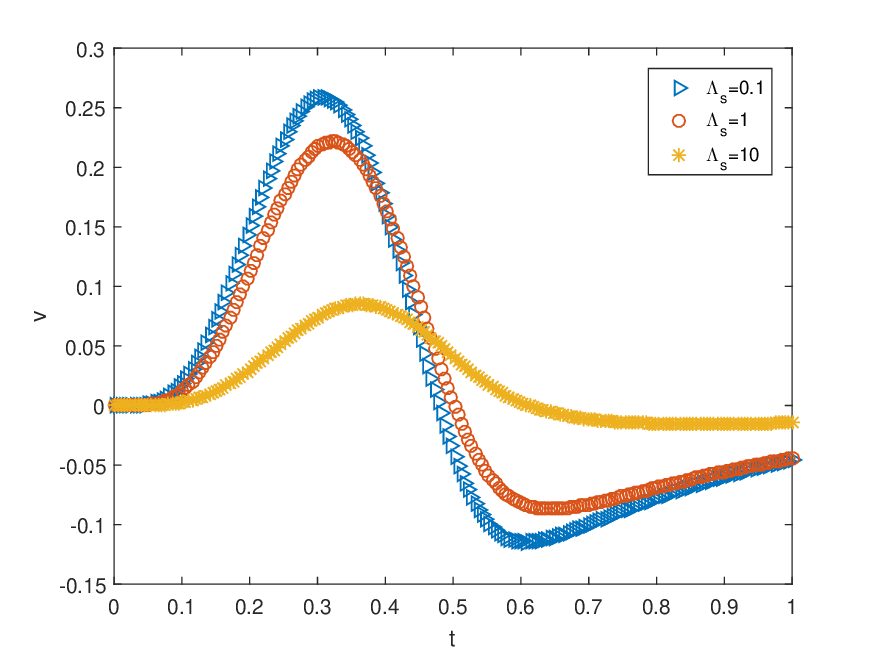}
 	\caption{Evolution of the slip velocity $v(0,t)$ in Case 3 (single plate oscillation) with $\theta =0.5$ and $\varLambda_s=0.5$ for $B=0.1$ (strong slip), 1 (moderate slip), and 10 (weak slip).} \label{fig:a8}
 \end{figure}

Figure \ref{fig:a10} shows the effect of the relaxation parameter  $\varLambda_s$ on the slip velocity $v(0,t)$  for $B=1$ (moderate slip). As for the case of continuous harmonic boundary motion, the effect of the dynamic slip parameter becomes stronger as the boundary $y=0$ is approached as may be seen in Figure \ref{fig:a3}. The same remains valid for the effect of slip parameter $B$ on the fluid motion as shown in Figures \ref{fig:a5} and \ref{fig:a8}, where the value of the dynamic slip parameter was set $\varLambda_s = 0.5$.

\begin{figure}[htbp]
	\centering
	\includegraphics[width=0.6\textwidth]{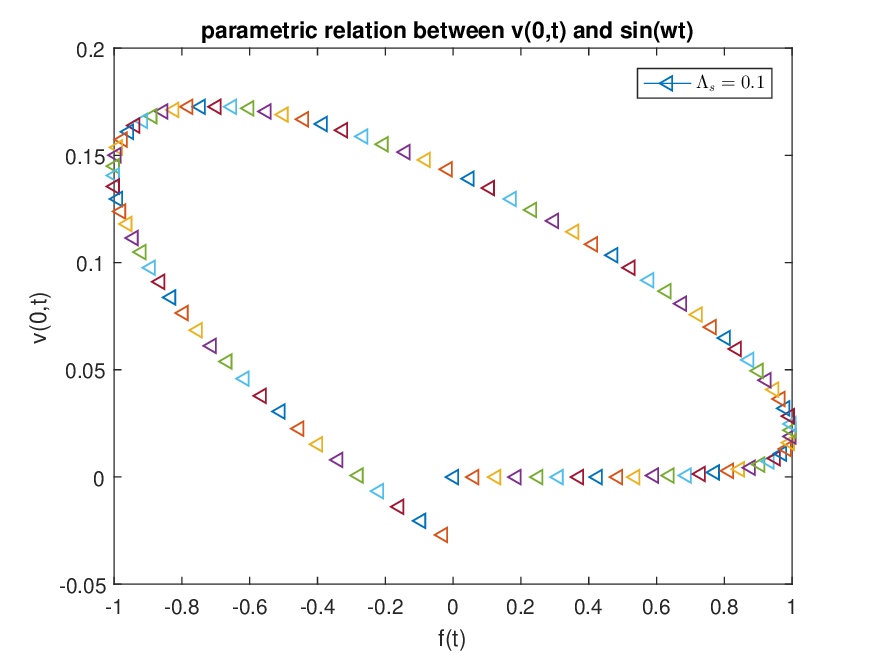}

{\bf (a)}

	\includegraphics[width=0.6\textwidth]{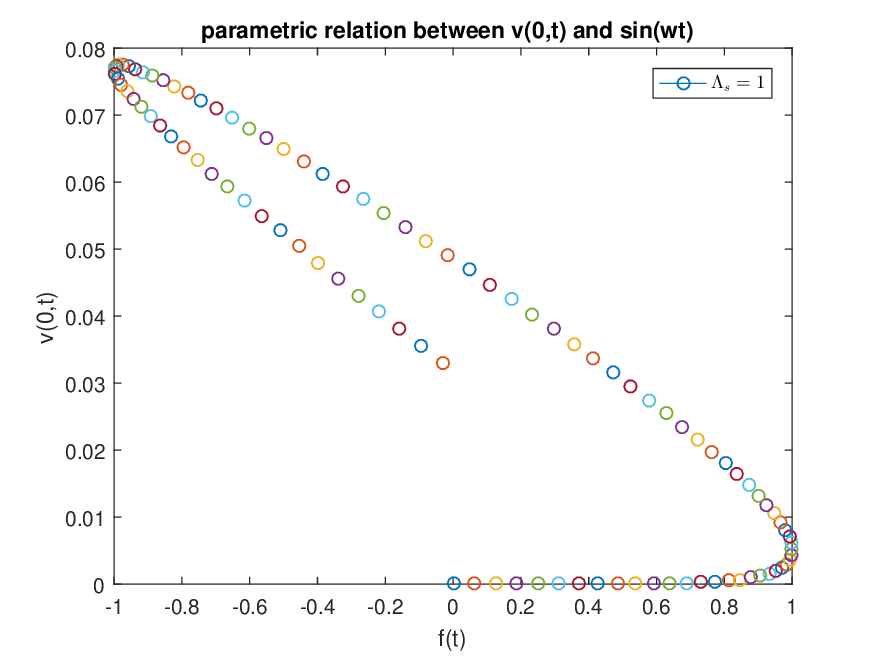}

{\bf (b)}

	\includegraphics[width=0.6\textwidth]{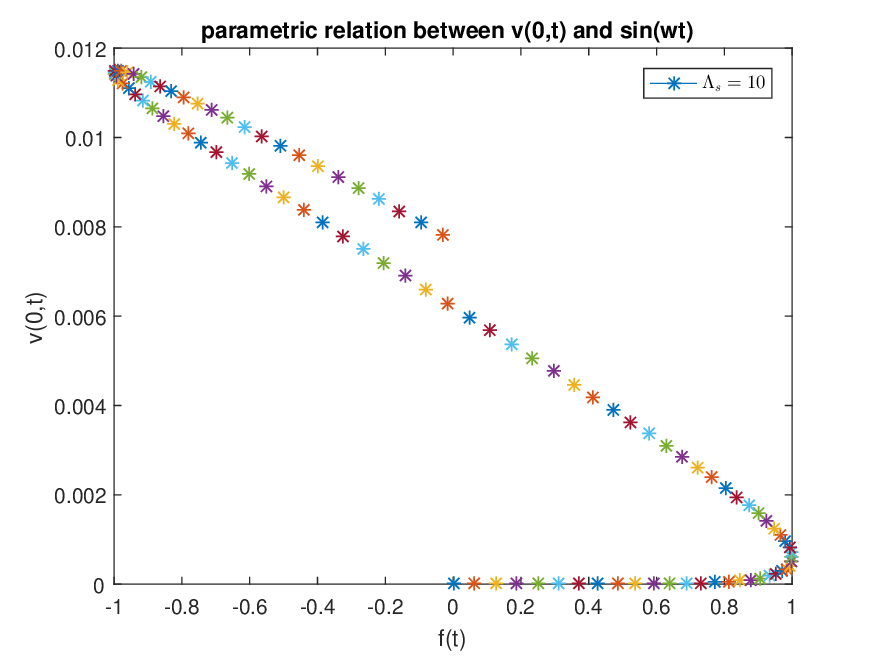}

{\bf (c)}

	\caption{Hysteretic behavior of the slip velocity $v(0,t)$ in Case 2 (single plate oscillation) with   $\theta =0$ and $B=1$: (a)   $\varLambda_s=0.1$; (b)  $\varLambda_s=1$; (c) $\varLambda_s=10$. Note that the scale of the vertical axis changes.} \label{fig:hyst-1}
\end{figure}

Figure \ref{fig:hyst-1} represents the hysteretic behavior of the slip velocity in following the motion of the wall in Case 3 for $B=1$ and three values of the dynamic slip parameter $\varLambda_s$. Here again, it is noticed that the hysteresis loop becomes narrower as the value of $\varLambda_s$ increases.

\begin{figure}[htbp]
	\centering
	\includegraphics[width=0.7\textwidth]{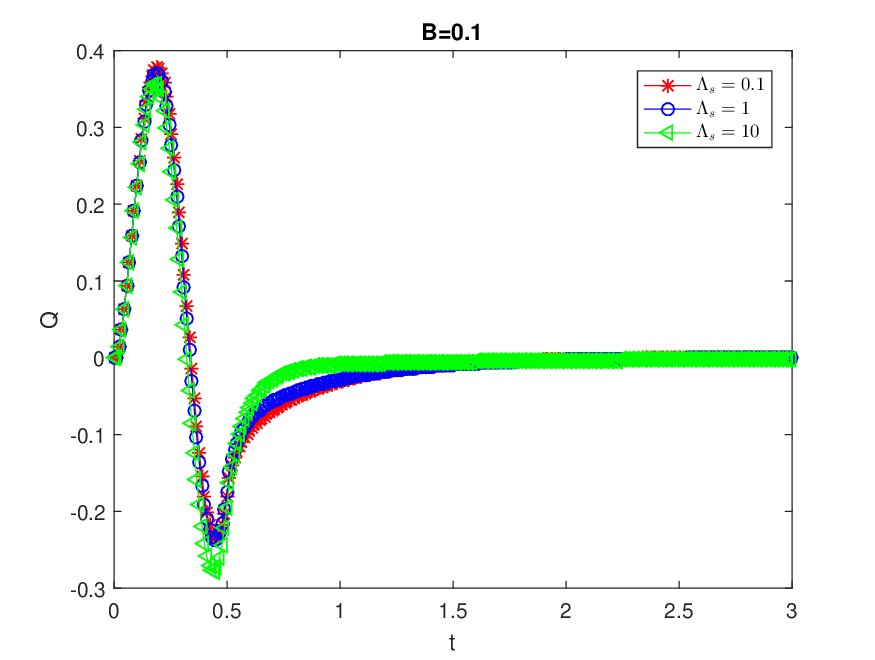}

{\bf (a)}

	\includegraphics[width=0.7\textwidth]{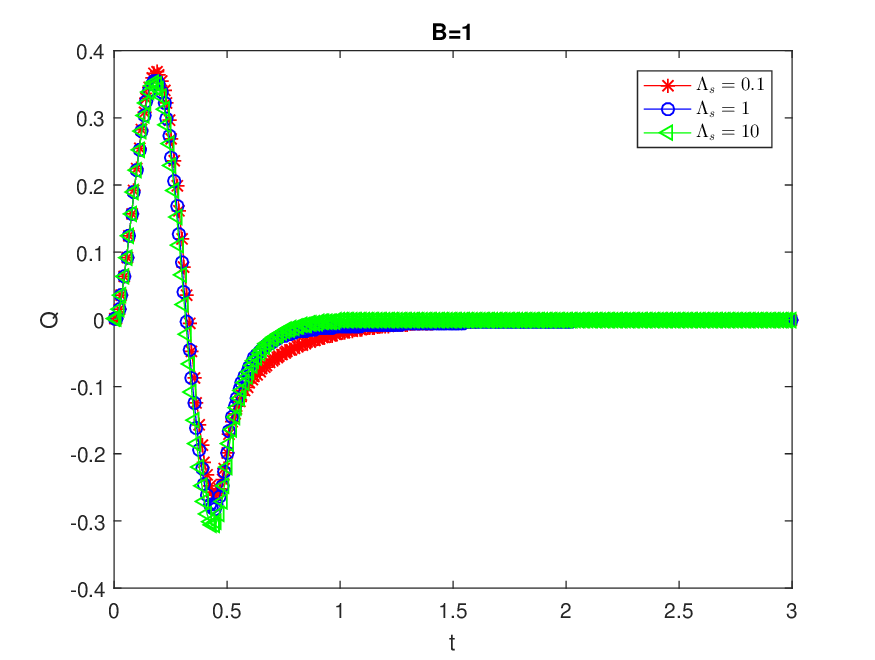}

{\bf (b)}

	\caption{Evolution of the volumetric flow rate $Q(t)$ in Case 3 (single plate oscillation)  with   $\theta =0$ and  $\varLambda_s=0.1$, 1, and 10: (a) $B=0.1$ (strong slip); (b) $B=1$ (moderate slip, right).}\label{fig:212}
\end{figure}

Finally, we have shown in Figure \ref{fig:212} the volumetric flow rate in Case 3 with $\theta = 0$, for two values of $B$ and three values of $\varLambda_s$. The volumetric flow rate is almost independent of these two slip parameters.

\section{Conclusions}
A weighted average finite difference scheme has been used to solve the time-dependent (start-up) plane Couette flow with dynamic slip along the fixed plate. Three different cases have been considered for the motion of the moving plate, i.e., constant-speed, sinusoidal and single-oscillation. The numerical solutions compare well with available analytical and numerical solutions. Both the slip and relaxation parameters appear to decelerate the evolution of the flow. For the two cases of accelerated (sinusoidal and one-period sinusoidal) boundary motion, the numerical results have clearly demonstrated a hysteretic behavior of the slip velocity that will be responsible for time lag and loss of energy transfer by the moving wall to the fluid.

Future work will be devoted to extend the  numerical scheme for solving non-linear extensions of the problem studied here.
These include the non-Newtonian (e.g., power-law) flow under both Navier and dynamic slip at the fixed wall and the Newtonian flow
with slip obeying a non-linear dynamic slip law.

\section*{Conflict of interest} The authors declare that they have no
conflict of interest.
\section*{Data availability}
All data generated or analyzed during this study are included in this article
\bibliographystyle{alpha}\fontsize{10}{10}
	
\end{document}